\documentclass[pra,epsfig,twocolumn,floats,showpacs]{revtex4}
\usepackage{amssymb}
\usepackage{amsmath}
\usepackage{colordvi}
\usepackage{amsthm}
\usepackage{epsfig}
\usepackage{graphicx}

\begin{document}

\title{Non-monogamy of Quantum Discord and Upper Bounds for Quantum
Correlation}
\author{Xi-Jun Ren}
\email{renxijun@mail.ustc.edu.cn}
\affiliation{School of Physics and Electronics, Henan University, Kaifeng 475001, China}
\author{Heng Fan}
\email{hfan@iphy.ac.cn}
\affiliation{Beijing National Laboratory for Condensed Matter Physics, Institute of
Physics, Chinese Academy of Sciences, Beijing 100190, China}

\begin{abstract}
We consider a monogamy inequality of quantum discord in a pure tripartite
state and show that it is equivalent to an inequality between quantum
mutual information and entanglement of formation of two parties. Since this
inequality does not hold for arbitrary bipartite states, quantum
discord can generally be both monogamous and polygamous. We also carry out
numerical calculations for some special states. The upper bounds of quantum
discord and classical correlation are also discussed and we give physical
analysis on the invalidness of a previous conjectured upper bound of quantum
correlation. Our results provide new insights for further understanding of
distributions of quantum correlations.
\end{abstract}

\maketitle

\section{Introduction}

Quantum states possess quantum correlations which are classically
unobtainable and act as an invaluable resource for quantum information
processing. For a long time, interests on quantum correlations focused on
quantum entanglement which is a special kind of quantum correlation enabling
fascinating tasks such as quantum key distribution, quantum teleportation
and superdense coding, etc \cite{horodecki}. Quantum entanglement does not
exist in separable states which are mixtures of separable direct product
states \cite{werner}. However, recent researches show that some separable
quantum states can exhibit their quantumness in many interesting
circumstances. In Ref.\cite{knill}, Knill and Laflamme introduced an
interesting computation model, deterministic quantum computation with one
quantum bit (DQC1), for which unentangled states can
provide exponential speed up over the best known classical algorithms.
Together with some other interesting tasks, such as locking of large amount
of classical correlations with small classical communication in unentangled
states, they ignite interests and studies on more general nonclassical
correlations or quantumness of quantum states \cite%
{henderson,ollivier,cui,modi,winterprl,SKB,luo}. From an extensive
background, the nonclassical correlations or quantumness of quantum states
are always of fundamental importance for both quantum information theory and quantum
mechanics .

Among the nonclassical correlation measures proposed with different
motivations, quantum discord is an important one for capturing all the
nonclassical correlations in a bipartite quantum state \cite%
{ollivier,henderson}. Researches on quantum discord develop quickly in
recent years. Direct calculations were carried out for some interesting quantum
states \cite{datta,giorda}. Operational meanings of quantum discord were
given in terms of some other important concepts like quantum state merging \cite{SKB,shabani,cavalcanti}%
. The dynamics of quantum discord were discussed in \cite{auccaise,streltsov}%
. Especially, experiments for quantum discord were also carried out \cite%
{auccaise,auccaise1}. In this paper, we are mainly concerned with the
monogamy property of quantum discord.

Unlike the arbitrary shareability of classical correlations among
multipartite systems, the shareability of quantum correlations is always
constrained by some monogamy relation as in case of entanglement measure \cite%
{coffman}. It says that for a multipartite state $\rho _{A_{0}A_{1}...A_{n}}$
and a quantum correlation measure $E$, the quantum correlation between $%
A_{0} $ and $A_{1}$, $A_{2}$,...,$A_{n}$ as a whole should be larger than
the sum of correlations between $A_{0}$ and $A_{1}$, $A_{2}$,...,$A_{n}$
separately, i. e. $E_{A_{0}|A_{1}...A_{n}}\geq \sum_{i}E_{A_{0}A_{i}}$. The
underlying intuition is that their difference should be genuine multipartite
quantum correlations which may exist only among three or more parties. In
\cite{coffman,osborne}, the authors constructed the monogamy relation for
qubit systems and concurrence which is a entanglement measure first
introduced by Hill and Wootters \cite{hill}. The monogamy relation in
continuous systems was provided in \cite{hiroshima}. More discussions on
monogamy of different quantum correlation measures can be found in, for
example, Refs. \cite{monogamy1,monogamy2,fan1,fan2}. Since quantum discord
quantifies the quantum correlations in a bipartite state, it is interesting
to study whether it also respects monogamy relation. Recently, Prabhu
\textit{et al.} \cite{prabhu} and Giorgi \cite{giorgi} have studied the
following monogamy relation for quantum discord,%
\begin{equation}
D^{\leftarrow }(\rho _{AB})+D^{\leftarrow }(\rho _{AC})\leq D^{\leftarrow
}(\rho _{A|BC}).  \label{monogamy1}
\end{equation}%
They showed that such a monogamy relation generally does not hold. In this
paper, we will study a different kind of monogamy relation for quantum
discord,
\begin{equation}
D^{\rightarrow }(\rho _{AB})+D^{\rightarrow }(\rho _{AC})\leq D^{\rightarrow
}(\rho _{A|BC}),  \label{monogamy2}
\end{equation}%
for a pure tripartite state $\left\vert \Psi _{ABC}\right\rangle $. Because
of the asymmetry of quantum discord, the above two monogamy relations are
quite different. Physically, the inequality (\ref{monogamy1}) means that the
measurement is taken on two parties, $B$ and $C$, coherently in right hand
side of the inequality and individually in left hand side of the inequality.
However, the inequality (\ref{monogamy2}) means that only one local
measurement on party $A$ is performed.

The outline of this paper is as follows. In the following section, after
reviewing the definition of quantum discord, we derive an equivalent
relation to the monogamy inequality (\ref{monogamy2}). Then through concrete
examples we numerically show that the monogamy relation (\ref{monogamy2}) does not
generally hold. With squashed entanglement we also provide a special case
when monogamy relation (\ref{monogamy2}) does hold. In the third section, we
discuss another interesting issue on quantum discord, the upper
bounds of quantum and classical correlations. Here, we give physical
explanations on the invalidity of previous conjectured upper bound of
quantum correlations. Finally we give our conclusions.

\section{Monogamy relations of quantum discord in a pure tripartite state}

In this section we will point out that the monogamy inequality (\ref%
{monogamy2}) can be reduced to a relation between entanglement of
formation (EoF) \cite{bennett}, a well-accepted entanglement measure, and
quantum mutual information. Before expanding our discussions, we first
review the definition of quantum discord. The definition is based on the difference between the total correlation and the classical correlation in the state, quantified by quantum mutual information and quantum conditional entropy by a local measurement, respectively. 
For a general bipartite state $\rho _{AB}$%
, quantum mutual information $I(\rho _{AB})=S(\rho _{A})+S(\rho _{B})-S(\rho
_{AB})$, is generally taken to be the measure of total correlations, both
classical and quantum. In order to quantify the classical correlation, a
positive operator valued measurement (POVM) $\{\Pi _{i}\}$ is made on party $%
A$, the resulting state given by the shared ensemble $\{p_{i},\rho _{B|i}\}$%
, where $p_{i}=Tr_{A,B}(\Pi _{i}\rho _{AB}),\rho _{B|i}=Tr_{A}(\Pi _{i}\rho
_{AB})/p_{i}$. Similar to the classical conditional entropy, quantum
conditional entropy is defined as $S_{\{\Pi _{i}\}}(B|A)=\sum_{i}p_{i}S(\rho
_{B|i})$, then an alternative version of quantum mutual information with
respect to POVM $\{\Pi _{i}\}$ is defined as $J_{\{\Pi _{i}\}}^{\rightarrow
}(\rho _{AB})=S(\rho _{B})-S_{\{\Pi _{i}\}}(B|A)$. Maximizing $J_{\{\Pi
_{i}\}}^{\rightarrow }(\rho _{AB})$ over all POVMs $\{\Pi _{i}\}$, we arrive at a
measurement independent quantity $J^{\rightarrow }(\rho _{AB})=\max_{\{\Pi
_{i}\}}[S(\rho _{B})-S_{\{\Pi _{i}\}}(B|A)]$ which captures all the
classical correlation present in $\rho _{AB}$. Taking the difference between total correlations and classical correlation, we obtain the following one way quantum
discord,
\begin{eqnarray}
D^{\rightarrow }(\rho _{AB}) &=&I(\rho _{AB})-J^{\rightarrow }(\rho _{AB})
\notag \\
&=&S(\rho _{A})-S(\rho _{AB})+\min_{\{\Pi _{i}\}}\sum_{i}p_{i}S(\rho _{B|i}).
\label{discord}
\end{eqnarray}%
Symbol $\rightarrow $ shows that such defined correlation measure is
asymmetric, i.e. generally $D^{\rightarrow }(\rho _{AB})\neq D^{\leftarrow
}(\rho _{AB})$, where $D^{\leftarrow }(\rho _{AB})$ is based on POVM on
party $B$.

Now, let us consider a pure tripartite state $\left\vert \Psi
_{ABC}\right\rangle $. The quantum discord between $A$ and $BC$ as a whole
is the von Neumann entropy of $A$,
\begin{equation*}
D^{\rightarrow }(\rho _{A|BC})=S(\rho _{A}).
\end{equation*}%
This means that by a von Neumann measurement with basis in agreement with
the Schmidt decomposition of bipartite partition $\left\vert \Psi
_{A|BC}\right\rangle $, the result is the quantum discord. On the other
hand, for pure state $\left\vert \Psi _{ABC}\right\rangle $, we have the
following relations between quantum discord and EoF \cite{fanchini},
\begin{equation}
D^{\rightarrow }(\rho _{AB})=S(\rho _{A})-S(\rho _{AB})+E_{F}(\rho _{BC}),
\label{dab}
\end{equation}%
\begin{equation}
D^{\rightarrow }(\rho _{AC})=S(\rho _{A})-S(\rho _{AC})+E_{F}(\rho _{BC}).
\label{dac}
\end{equation}%
From these relations, we have%
\begin{eqnarray*}
&&D^{\rightarrow }(\rho _{A|BC})-D^{\rightarrow }(\rho _{AB})-D^{\rightarrow
}(\rho _{AC}) \\
&=&S(\rho _{AB})+S(\rho _{AC})-S(\rho _{A})-2E_{F}(\rho _{BC}) \\
&=&S(\rho _{B})+S(\rho _{C})-S(\rho _{BC})-2E_{F}(\rho _{BC}) \\
&=&I(\rho _{BC})-2E_{F}(\rho _{BC}).
\end{eqnarray*}%
Therefore the monogamy relation (\ref{monogamy2}) is reduced to
\begin{equation}
E_{F}(\rho _{BC})\leq \frac{I(\rho _{BC})}{2}.  \label{monogamy3}
\end{equation}%
Inequality (\ref{monogamy3}) shows that an inequality between quantum mutual
information and EoF of a bipartite state implies the monogamy inequality of
quantum discord in a tripartite state, which is the purification of the
bipartite state. In \cite{fanchi}, $I(\rho _{BC})-2E_{F}(\rho _{BC})$ was
also found to be equal to the difference of classical correlation $%
J^{\rightarrow }(\rho _{AB})$ and quantum discord $D^{\rightarrow }(\rho
_{AB})$ between $AB$ which was named discrepancy.\ We know that quantum
mutual information is commonly considered to quantify the total correlations
\cite{groisman} and entanglement of formation is a measure of entanglement.
Interestingly, for any pure bipartite state $|\Psi _{AB}\rangle $, we have
that quantum mutual information, $I(\Psi _{AB})=2S(\rho _{A})$, is two times
of the entanglement of formation of state $|\Psi _{AB}\rangle $. This
inequality seems reasonable. Actually inequality (\ref{monogamy3}) has
already been analyzed as a postulate for measures of quantum correlation in
\cite{li}, where Li and Luo show that there are states for which inequality (%
\ref{monogamy3}) does not hold and they also argue that EoF may not be a
proper quantum correlation measure consistent with quantum mutual
information. Here, the existence of states violating eq.(\ref{monogamy3})
shows that monogamy relation (\ref{monogamy2}) does not generally hold.
In the following we will discuss monogamy relation (\ref{monogamy2}) through
some tripartite states.

\emph{Generalized pure three-qubit GHZ and W states.}---In Ref. \cite{prabhu}%
, a necessary and sufficient condition for quantum discord being monogamous
with inequality(\ref{monogamy1}) is given and applied to generalized GHZ and
W states. Based on numerical calculations of generalized W state, it was
conjectured that the quantum discord in these states is polygamous which is
confirmed in Ref. \cite{giorgi} with the conservation law for distributed
EoF and quantum discord \cite{fanchini}. Here, concerning with our proposed
monogamy inequality (\ref{monogamy2}), it can be easily seen that for
generalized GHZ states $\left\vert \Psi _{ABC}^{GHZ}\right\rangle =\alpha
\left\vert 000\right\rangle _{ABC}+\beta \left\vert 111\right\rangle _{ABC}$%
, relation (\ref{monogamy3}) holds since EoF is simply zero and hence
monogamy relation (\ref{monogamy2}) holds. However, generalized W states are
different and direct calculations show that quantum discord in generalized W
states can be both monogamous and polygamous with inequality (\ref{monogamy2}%
). Explicitly, the generalized W states take the form,
\begin{equation}
\left\vert \Psi _{ABC}^{W}\right\rangle =\alpha \left\vert 011\right\rangle
_{ABC}+\beta \left\vert 101\right\rangle _{ABC}+\gamma \left\vert
110\right\rangle _{ABC}.  \label{wstate}
\end{equation}%
Without lose of generality, we assume that $\alpha ,\beta ,\gamma $ are all
real, and the normalization condition $\alpha ^{2}+\beta ^{2}+\gamma ^{2}=1$%
. We calculate $I(\rho _{BC})-2E_{F}(\rho _{BC})$ for this state and the
results are given in Fig.1. It can be easily seen that though in most cases
quantum discord is polygamous ($D_{A\rightarrow BC}-D_{A\rightarrow
B}-D_{A\rightarrow C}<0$), when $\beta ^{2},\gamma ^{2}$ are small we have $%
D_{A\rightarrow BC}-D_{A\rightarrow B}-D_{A\rightarrow C}\geq 0$, which
means that quantum discord is monogamous between $A$ and $B,C$. In the
reduced bipartite state $\rho _{BC}$, $\alpha ^{2}$ is the proportion of
direct product state $\left\vert 11\right\rangle $, $\beta ^{2}+\gamma ^{2}$
quantifies the proportion of entangled state $\beta \left\vert
01\right\rangle _{BC}+\gamma \left\vert 10\right\rangle _{BC}$. Fig.1
implies that monogamy relation (\ref{monogamy2}) is roughly related to the
entangled proportion $\beta ^{2}+\gamma ^{2}$ shared between $B$ and $C$.
Monogamy holds when this entangled proportion is small and polygamy holds
when it is large. Similar analysis can be made on generalized GHZ class
states which contain the above generalized GHZ states $\left\vert \Psi
_{ABC}^{GHZ}\right\rangle $ as a subset. One simple example is, $\left\vert
\Psi _{ABC}\right\rangle =\alpha \left\vert 000\right\rangle _{ABC}+\beta
\left\vert 100\right\rangle _{ABC}+\gamma \left\vert 111\right\rangle _{ABC}$%
, one can find that these states can also be both monogamous and polygamous.

\begin{figure}[tbp]
\includegraphics[width=0.38\textwidth]{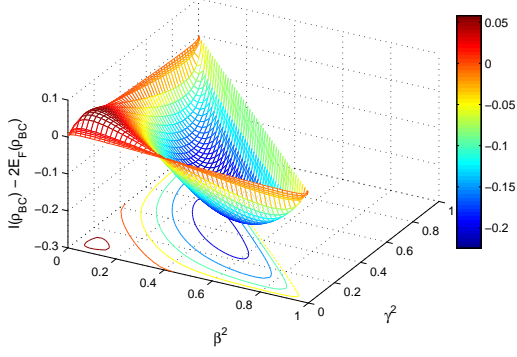}
\caption{Difference between mutual information and two times EoF for a
bipartite state of parties $B$ and $C$ from a generalized $W$ state in Eq.(%
\protect\ref{wstate}). It can be found that quantum discord can be both
monogamous and polygamous.}
\end{figure}

\begin{figure}[tbp]
\includegraphics[width=0.38\textwidth]{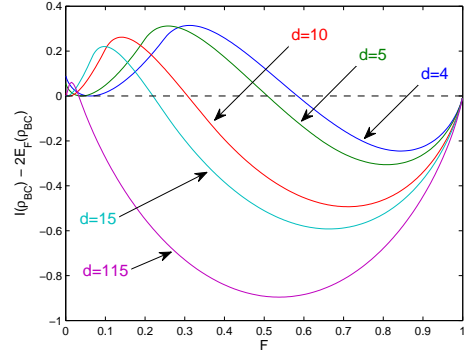}
\caption{Difference between mutual information and two times EoF for
isotropic state in Eq.(\protect\ref{isotropicstate}). When dimension $d$
increases, the region of negative becomes larger corresponding to polygamy
for a pure tripartite state.}
\end{figure}

\emph{Isotropic states and Werner states.}---Since monogamy inequality (\ref%
{monogamy2}) is equivalent to inequality (\ref{monogamy3}) which concerns
only about a bipartite state $\rho _{BC}$, while party $A$ can be regarded
as an extension of this bipartite state for purification, we need only to
analyze bipartite state $\rho _{BC}.$ In the following we will consider
bipartite isotropic states \cite{Terhal} and Werner states \cite{vollbrecht}
which have analytical expressions for EoF \cite{Terhal}. The isotropic
states take the following form,
\begin{equation}
\rho _{BC}=\frac{1-F}{d^{2}-1}(I-\left\vert \Psi ^{+}\right\rangle
\left\langle \Psi ^{+}\right\vert )+F\left\vert \Psi ^{+}\right\rangle
\left\langle \Psi ^{+}\right\vert  \label{isotropicstate}
\end{equation}%
where, $0\leq F\leq 1$ and $\left\vert \Psi ^{+}\right\rangle =\frac{1}{%
\sqrt{d}}\sum_{i=1}^{d}\left\vert ii\right\rangle $, $d$ is dimension of
Hilbert space $B$ and $C$. First, since both reduced density matrices of $%
B,C $ are $I/d$, we have $S(\rho _{B})=S(\rho _{C})=\log _{2}d$. Second, $%
S(\rho _{BC})$ can be directly calculated as,
\begin{eqnarray}
S(\rho _{BC}) &=&-F\log _{2}F-(1-F)\log _{2}(1-F)  \notag \\
&&+(1-F)\log _{2}(d^{2}-1).
\end{eqnarray}%
From Ref.\cite{Terhal}, we know that the FoF of $\rho _{BC}$ is,
\begin{equation*}
E_{F}(\rho _{BC})=\left\{
\begin{array}{l}
0,~F\in \mathrm{I} \\
H_{2}(\gamma (F))+(1-\gamma (F))\log (d-1), \\
~~~~~~~~F\in \mathrm{II} \\
\frac{d\log (d-1)}{d-2}(F-1)+\log d,F\in \mathrm{III}%
\end{array}%
\right.
\end{equation*}%
where cases I,II,III are $[0,\frac{1}{d}],[\frac{1}{d},\frac{4(d-1)}{d^{2}}%
],[\frac{4(d-1)}{d^{2}},1]$, respectively, $\gamma (F)=\frac{1}{d}\left(
\sqrt{F}+\sqrt{(d-1)(1-F)}\right) ^{2}$, $H_{2}(x)=-x\log _{2}x-(1-x)\log
_{2}(1-x)$.

In Fig.2, we plot $I(\rho _{BC})-2E_{F}(\rho _{BC})$ for $d=4,5,10,15,115$,
whose lines are arranged from right to left. It can be seen that quantum
discord is monogamous when $F$ is small which means $\rho _{BC}$ has less
singlet fractions. With the increasing of dimension $d$, singlet $\left\vert
\Psi ^{+}\right\rangle $ has higher proportion and the quantum discord has
larger polygamous region. These results are consistent with the results
obtained for generalized pure three-qubit GHZ and W states.

Completely similar analysis can be made on the following Werner states,%
\begin{equation}
w_{BC}(x)=\frac{d-x}{d^{3}-d}I+\frac{dx-1}{d^{3}-d}P,x\in \lbrack -1,1]
\label{wernerstate}
\end{equation}%
where $P=\sum_{i,j=1}^{d}\left\vert ij\right\rangle \left\langle
ji\right\vert $ is the flip operator. We note that these analysis has been
carried out in \cite{li} and the results are quite similar with isotropic
states given above.

\emph{Observation based on squashed entanglement.}---For a pure tripartite
state $\left\vert \Psi _{ABC}\right\rangle $, if its reduced bipartite state
$\rho _{BC}$ satisfies $E_{D}(\rho _{BC})=E_{F}(\rho _{BC})$, then the
monogamy relations (\ref{monogamy2},\ref{monogamy3}) hold, where $E_{D}(\rho
_{BC})$ is the entanglement of distillation. This observation can be proved
with squashed entanglement which is defined in terms of conditional mutual
information \cite{christandl},%
\begin{equation}
E_{sq}(\rho _{BC}):=\inf \{\frac{1}{2}I(B,C|E):\rho _{BCE}\text{ extension
of }\rho _{BC}\}.  \label{squashedentanglement}
\end{equation}%
In \cite{christandl}, it has been proved that $E_{D}\leq E_{sq}(\rho _{BC})$%
. Meanwhile we have $E_{sq}(\rho _{BC})\leq \frac{1}{2}I(\rho _{BC})$ since $%
I(B,C|E)$ is the \textquotedblleft squashed\textquotedblright\ correlation
from $I(\rho _{BC})$ where the classical correlations are squashed out as
much as possible. Obviously, when $E_{D}=E_{F}$, inequality (\ref{monogamy3}%
) is satisfied and quantum discord monogamy relation (\ref{monogamy2})
between $A$ and $B,C$ in $\left\vert \Psi _{ABC}\right\rangle $\ holds.

\section{upper bounds on quantum discord and classical correlations}

In Ref.\cite{luo1}, it was conjectured that, given a bipartite sate $\rho
_{AB}$ defined in the Hilbert space $H_{A}\otimes H_{B}$, the following
upper bounds for quantum discord and classical correlations could exist:%
\begin{eqnarray}
J^{\rightarrow }(\rho _{AB}) &\leq &\min [S(\rho _{A}),S(\rho _{B})],
\label{classcialupper} \\
D^{\rightarrow }(\rho _{AB}) &\leq &\min [S(\rho _{A}),S(\rho _{B})],
\label{quantumupper}
\end{eqnarray}%
In \cite{zhang}, the upper bound of classical correlation (%
\ref{classcialupper}) is proved to be true and the upper bound of quantum correlation (%
\ref{quantumupper}) is proved to be true only under some conditions. In \cite{giorgi},
based on known results of three qubits \cite{coffman}, Giorgi showed that
the above two upper bounds hold for rank-2 states of two qubits.
Now, with the features of quantum discord, we provide some more concise
discussions on these upper bounds. First, we assume the mixed $\rho _{AB}$ to be reduced from a pure tripartite state $\left\vert \phi
_{ABC}\right\rangle $ such that $Tr_{C}\left\vert \phi _{ABC}\right\rangle
\left\langle \phi _{ABC}\right\vert =\rho _{AB}$.

\emph{Upper bound on classical correlation.}---According to its definition,
the classical correlation is $J^{\rightarrow }(\rho _{AB})=S(\rho _{B})-\min
S(B|\{E_{j}^{A}\})=S(\rho _{B})-E_{F}(\rho _{BC})\leq S(\rho _{B})$, the
last inequality comes from the fact $E_{F}(BC)\geq 0$. In order to show that
we simultaneously have $J^{\rightarrow }\leq S(\rho _{A})$, we need to prove
the following inequality,%
\begin{equation}
E_{F}(\rho _{BC})\geq S(\rho _{B})-S(\rho _{BC}).
\label{coherentinformation}
\end{equation}%
This obviously holds since $S(\rho _{B})-S(\rho _{BC})$ is the coherent
information \cite{horodecki1,devetak} which is a lower bound for distillable
entanglement smaller than EoF. Therefore we know that the upper bounds for
classical correlation in (\ref{classcialupper}) holds.

\emph{Upper bound on quantum correlation.}---Here the quantum correlation
measure is just quantum discord, it is $D^{\rightarrow }(\rho _{AB})=S(\rho
_{A})-S(\rho _{AB})+\min S(B|\{E_{j}^{A}\})=S(\rho _{A})-S(\rho
_{C})+E_{F}(\rho _{BC})=S(\rho _{A})-S(\rho _{C})+\min
S(C|\{E_{j}^{A}\})=S(\rho _{A})-J^{\rightarrow }(\rho _{AC})\leq S(\rho
_{A}) $, the last inequality comes from the fact that $J^{\rightarrow }(\rho
_{AC})\geq 0$ or the concavity of entropy if we consider that $S(\rho
_{C})-E_{F}(\rho _{BC})\geq 0$. With one half of inequality (\ref%
{quantumupper}) proved, can we simultaneously prove another half of the
inequality, $D^{\rightarrow }(\rho _{AB})\leq S(\rho _{B})$ ? We only need
to consider the case $S(\rho _{A})>S(\rho _{B})$, then we should have,%
\begin{equation}
E_{F}(\rho _{BC})\leq S(\rho _{B})+S(\rho _{AB})-S(\rho _{A}),
\label{condition1}
\end{equation}%
Since $\left\vert \phi _{ABC}\right\rangle $ is a pure state, it is
equivalent to
\begin{equation}
E_{F}(\rho _{BC})\leq I(\rho _{BC}).  \label{condition3}
\end{equation}%
In general we consider that mutual information quantifies the total
correlations, it naturally seems to be larger than EoF which only quantifies
quantum correlation. However this is not true! Hayden, Leung and Winter \cite%
{hayden} found that EoF in a bipartite state can be larger than its mutual
information. In \cite{li}, Li and Luo consolidated their findings and showed
that Werner state \cite{vollbrecht} has this property. From a Werner state $%
\rho _{BC}$ (\ref{wernerstate}) which violates (\ref{condition3}), a
purified tripartite state $\left\vert \phi _{ABC}\right\rangle $ can be
constructed with $S(\rho _{A})>S(\rho _{B})$. Then its reduced bipartite
state $\rho _{AB}$ violates the upper bound in (\ref{quantumupper}). On the
other hand, the violation of this upper bound can be understood from the viewpoint of
coherent information. Since $E_{F}(\rho _{BC})+J^{\rightarrow }(\rho
_{AB})=S(\rho _{B})$ \cite{monogamy1}, inequality (\ref{condition1}) is
equivalent to
\begin{equation}
S(\rho _{A})-S(\rho _{AB})\leq J^{\rightarrow }(\rho _{AB}),
\label{condition2}
\end{equation}%
where $S(\rho _{A})-S(\rho _{AB})$ is one-way coherent information with
classical communication from $\mathit{B}$ to $\mathit{A}$ and $%
J^{\rightarrow }(\rho _{AB})$ is one-way distillable common randomness with
classical communication from $\mathit{A}$ to $\mathit{B}$, see Ref. \cite%
{monogamy1}. We know that coherent information is a lower bound for
distillable entanglement which is a lower bound for secret key, while the
secret key rate between $\mathit{A}$ and $\mathit{B}$ is obviously smaller
than their distillable common randomness. Therefore the violation of (\ref%
{condition2}) means that there are states from which the distillable secret
key with classical communication from $\mathit{B}$ to $\mathit{A}$ can even
be larger than the distillable common randomness with classical
communication from $\mathit{A}$ to $\mathit{B}$.

Thus in general the conjectured upper bound of quantum discord in relation (%
\ref{quantumupper}) does not hold, however, a released upper bound of
quantum discord can be obtained,
\begin{eqnarray}
D^{\rightarrow }(\rho _{AB}) &\leq & \max [S(\rho _{A}),S(\rho _{B})].  \label{proper}
\end{eqnarray}

\section{Conclusion}

Quantum discord is an important quantum correlation measure. In this paper,
we discuss a monogamy relation for this measure which is different from the
monogamy relation proposed in Refs.\cite{prabhu,giorgi}. For a tripartite
pure state, monogamy relation (\ref{monogamy2}) is reduced to a relation (%
\ref{monogamy3}) between mutual information and entanglement of formation in
a reduced bipartite state. Since relation (\ref{monogamy3}) does not
generally hold, monogamy relation (\ref{monogamy2}) can be both monogamous
and polygamous for arbitrary pure tripartite states. Through numerical calculations with several explicit classes of
states, we show that monogamy relation (\ref{monogamy2}) is roughly related
to the entangled proportion shared in the reduced bipartite state. It holds
when the entangled proportion is small and turns into a polygamy relation
when the entangled proportion is large.

In this article, we also provide a concise discussion on a conjecture of
upper bounds for classical and quantum correlations in a bipartite state
\cite{luo}. We show that the upper bounds on quantum correlation may be
violated with a pure tripartite state constructed from a Werner state. The
physical meaning behind the violation is discussed with operational quantum information concepts. At the same time, a released upper bound (\ref{proper}) still holds. Our results should be useful for
further understanding of quantum discord and distribution of classical and
quantum correlations in multipartite states.

\noindent \emph{Acknowledgments.}---We thank Shuai Cui for useful
discussions. This work is supported by \textquotedblleft
973\textquotedblright\ program (2010CB922904) and NSFC (10974247, 11175248, U1204114,
11047174). X-J Ren acknowledges financial support from the education
department of Henan province.

\end{document}